\documentstyle[preprint,aps,prbbib]{revtex}                                                 
\begin{document}     
\begin{center}{\Large\bf Value of the Cosmological Constant in the
Cosmological Relativity Theory}\end{center}
\begin{center}{Moshe Carmeli and Tanya Kuzmenko}\end{center}
\begin{center}{Department of Physics, Ben Gurion University, Beer Sheva 84105, 
Israel}\end{center}
\begin{center}{Email: carmelim@bgumail.bgu.ac.il}\end{center}
\begin{abstract}
It is shown that the cosmological relativity theory predicts the
value $\Lambda=2.28\times 10^{-35}s^{-2}$ for the cosmological constant. This 
value of $\Lambda$ is in excellent agreement with the measurements recently
obtained by the {\it High-Z Supernova Team} and the {\it Supernova Cosmology 
Project}.
\end{abstract}
\newpage
\section{Introduction}
The problem of the cosmological constant and the vacuum energy associated with
it is of high interest these days. There are many questions related to it at
the quantum level, all of which are related to quantum gravity. Why there 
exists the critical mass density and why the cosmological constant has this
value? Trying to answer these questions and others were recently the subject 
of many publications [1-18].

In this paper it is shown that the cosmological relativity 
theory [19] predicts the value $\Lambda=2.28\times 10^{-35}$s$^{-2}$ for the 
cosmological constant. This value of $\Lambda$ is in excellent agreement with
the measurements recently obtained by the {\it High-Z Supernova Team} and the
{\it Supernova Cosmological Project} [20-26].
\section{The Cosmological Constant}
In 1922 Friedmann solved the Einstein gravitational field equations and 
obtained nonstatic cosmological solutions presenting an expanding universe [27].
Einstein, who thought at that time that the universe should be static and 
unchanged forever, suggested a modification to his original field equations
by adding to them the so-called cosmological term which can stop the 
expansion. The field equations with the added term are:
$$R_{\mu\nu}-\frac{1}{2}g_{\mu\nu}R+\Lambda g_{\mu\nu}=\kappa T_{\mu\nu},
\eqno(1),$$
where $\Lambda$ is the cosmological constant, the value of which is supposed to
be determined by experiment. In Eq. (1) $R_{\mu\nu}$ and $R$ are the Ricci 
tensor and scalar, respectively, $\kappa=8\pi G$, where $G$ is Newton's constant
and the speed of light is taken as unity.

Soon after that Hubble [28] found experimentally that the distant galaxies are
receding from us, and the farther the galaxy the bigger its velocity as
determined by its redshift.

After Hubble's discovery that the universe is expanding, the role of the
cosmological constant to allow static homogeneous solutions to Einstein's
equations in the presence of matter, was looked upon as unnecessary. For a
long time the cosmological term was considered to be of no interest in
cosmological physical problems. 
\section{The Friedmann Universe} 
For a homogeneous and isotropic universe with the line element [29,30]
$$ds^2=dt^2-a^2(t)R_0^2\left[\frac{dr^2}{1-kr^2}+r^2\left(d\theta^2+\sin^2
\theta d\phi^2\right)\right],\eqno(2)$$
where $k$ is the curvature parameter ($k=1,0,-1$) and $a(t)=R(t)/R_0$ is the
scale factor, with the energy-momentum tensor
$$T_{\mu\nu}=(\rho+p)u_\mu u_\nu-pg_{\mu\nu},\eqno(3)$$
Einstein's equations (1) reduce to the two Friedmann equations
$$H^2\equiv\left(\frac{\dot{a}}{a}\right)^2=\frac{\kappa}{3}\rho+
\frac{\Lambda}{3}-\frac{k}{a^2R_0^2},\eqno(4)$$
$$\frac{\ddot{a}}{a}=-\frac{\kappa}{6}\left(\rho+3p\right)+\frac{\Lambda}{3}.
\eqno(5)$$  
In Eqs. (4) and (5) $H$ is Hubble's parameter, $\rho$ is the mass density and
$p$ is the pressure. These equations admit a static solution ($\dot{a}=0$) 
with $k>0$ and $\Lambda>0$.

From the Friedmann equation (4) it then follows that for any value of the 
Hubble parameter $H$
there exists a critical mass density $\rho_c=3H_0^2/\kappa$ at which the 
spatial geometry is flat ($k=0$). One usually measures the
total mass density in terms of the critical density $\rho_c$ by means of the
density parameter $\Omega=\rho/\rho_c$.

In general, the mass density $\rho$ includes contributions from various
distinct components. From the point of view of cosmology, the relevant aspect
of each component is how its contribution to the total energy density evolves 
as the universe expands. A positive $\Lambda$ causes acceleration
to the universe expansion, whereas a negative $\Lambda$ and ordinary matter
tend to decelerate it. Moreover, the relative contributions of the components
to the energy density change with time. For $\Omega_\Lambda<0$, the universe 
will always recollapse to a Big Crunch. For $\Omega_\Lambda>0$ the universe
will expand forever unless there is sufficient matter to cause recollapse
before $\Omega_\Lambda$ becomes dynamically important. For $\Omega_\Lambda=0$
we have the familiar situation in which $0<\Omega_M\leq 1$ universes expand
forever and $\Omega_M>1$ universes recollapse. (For more details see Ref. 19.)
\section{The Supernovae Experiments}
Recently two groups (the Supernovae Cosmology Project and the High-Z Supernova
Team) presented evidence that the expansion of the universe is accelerating 
[20-26]. These teams have measured the distances to cosmological supernovae by
using the fact that the intrinsic luminosity of Type Ia supernovae is closely
correlated to their decline rate from maximum brightness, which can be 
independently measured. These measurements, combined with redshift data for 
the supernovae, led to the prediction of an accelerating universe. Both teams
obtained
$$\Omega_M\approx 0.3,\hspace{5mm} \Omega_\Lambda\approx 0.7,\eqno(6)$$
and strongly ruled out the traditional ($\Omega_M$, $\Omega_\Lambda$)=(1, 0)
universe. This value of the density parameter $\Omega_\Lambda$ corresponds to
a cosmological constant that is small but, nevertheless, nonzero and positive,
$$\Lambda\approx 10^{-52}\mbox{\rm m}^{-2}\approx 10^{-35}\mbox{\rm s}^{-2}.
\eqno(7)$$
\section{The Cosmological Relativity Theory}
In Ref. 19 a four-dimensional cosmological
relativity theory that unifies space and velocity was presented. The theory
predicts that the universe accelerates and hence it is equivalent to having
a positive value for $\Lambda$ in it. As is well known, in the traditional 
work of Friedmann when added to it a cosmological constant, the field 
equations obtained are highly complicated and no solutions were obtained so
far.

Cosmological relativity theory, on the other hand, yields exact solutions and describes 
the universe as having a three-phase evolution with a decelerating expansion
followed by a constant and an accelerating expansion, and it predicts that the
universe is now in the latter phase. In the framework of this theory the 
zero-zero component of Einstein's equations is written as
$$R_0^0-\frac{1}{2}\delta_0^0R=\kappa\rho_{eff}=\kappa\left(\rho-\rho_c
\right),\eqno(8)$$
where $\rho_c=3/\kappa\tau^2\approx 3H_0^2/\kappa$ is the critical mass density
and $\tau$ is Hubble's time in the zero-gravity limit.

Comparing Eq. (8) with the zero-zero componentof Eq. (1), one obtains the
expression for the cosmological constant in cosmological relativity theory,
$$\Lambda=\kappa\rho_c=3/\tau^2\approx 3H_0^2.\eqno(9)$$
Assuming that Hubble's constant $H_0=85$km/s-Mpc, then 
$$\Lambda=2.28\times 10^{-35}\mbox{\rm s}^{-2}.\eqno(10)$$ 
This result is in excellent agreement with the recent 
supernovae experimental results.
\section{Conclusions}
We have seen how the cosmological constant can be determined in a natural way
without even adding it explicitely to the Einstein field equations. Rather,
the introduction of the effective mass density $\rho_{eff}=\rho-\rho_c$ is 
enough to ensure that the universe expands in the same way using 
Einstein's
field equations with a cosmological constant. But there is a big difference 
now: The theory determines the numerical value of the cosmological constant,
and experiments confirm it.

\end{document}